\documentclass[preprint,showpacs,preprintnumbers,amsmath,amssymb]{revtex4}

\usepackage{graphicx}
\usepackage{dcolumn}
\usepackage{bm}
\begin{document}

\title{Pressure effects in hollow and solid iron oxide nanoparticles}

\author{N. J. O. Silva$^1$, S. Saisho$^2$, M. Mito$^2$, A. Mill\'{a}n$^3$, F. Palacio$^3$, A. Cabot$^4$, \`{O}scar Iglesias$^5$, A. Labarta$^5$}
\address{$^1$Departamento de F\'{i}sica and CICECO, Universidade de Aveiro, 3810-193 Aveiro, Portugal}
\address{$^2$Faculty of Engineering, Kyushu Institute of
Technology, Kitakyushu, 804-8550, Japan}
\address{$^3$Instituto de Ciencia de Materiales de Arag\'{o}n,
CSIC - Universidad de Zaragoza. Departamento de F\'{i}sica de la
Materia Condensada, Facultad de Ciencias, 50009 Zaragoza, Spain.}
\address{$^4$ Universitat de Barcelona and Catalonia Energy Research Institute, Barcelona, Spain}
\address{$^5$ Departament de F\'{\i}sica Fonamental, Universitat de
Barcelona and Institut de Nanoci\`{e}ncia i Nanotecnologia,
Universitat de Barcelona, Mart\'{\i} i Franqu\`{e}s 1, 08028
Barcelona, Spain.}

\email{nunojoao@ua.pt}

\date{\today}

\begin{abstract}
We report a study on the pressure response of the anisotropy
energy of hollow and solid maghemite nanoparticles. The
differences between the maghemite samples are understood in terms
of size, magnetic anisotropy and shape of the particles. In
particular, the differences between hollow and solid samples are
due to the different shape of the nanoparticles and by comparing
both pressure responses it is possible to conclude that the shell
has a larger pressure response when compared to the core.
\end{abstract}

\maketitle

\section{\label{sec:Intro}Introduction}

Pressure is a thermodynamical parameter on which changes in
structural and magnetic properties are commonly observed. In the
case of nanoparticles (NPs), pressure modifies the transition
temperature\cite{ISI:000294484300134}, the susceptibility and
magnetization\cite{YukiPression, YukiJMMM}, the hysteresis
cycles\cite{YukiPression} and the effective anisotropy energy
barrier\cite{YukiAPL}. 
The effect of pressure in the NPs core and surface is distinct,
allowing disentanglement of core and shell magnetic properties.
This is the case of the anisotropy energy barrier in spherical
maghemite NPs; while at room pressure only an effective value can
be obtained, with increasing pressure core and shell size have a
different response and so does the effective anisotropy energy,
such that core and shell components of the anisotropy energy can
be extracted \cite{YukiAPL}.

Core/shell models have been successfully used in the context of
magnetic properties of maghemite NPs. These models often consider
that the particles are constituted by a bulk-like core and a
surface with distinct magnetic properties. In a seminal work, Coey
described maghemite NPs as having a core with the bulk spin
arrangement and a surface in which the spins are inclined at some
angle\cite{Coey}. The surface spins of maghemite NPs were lately
shown to have spin-glass like properties \cite{Martinez}. In the
case of magnetization ($M$) measurements, surface is often
considered as constituted by single paramagnetic and/or
aniferromagnetic ions, whose contribution to $M$ is linear in
field \cite{Angel_Ms}. The origin of this surface magnetic
behaviour is associated with incomplete coordination of
superficial ions and to the likely occurrence of structural
defects at the surface, as shown by experimental techniques and
computer simulations (see for instance Ref. \cite{Berkowitz_prl2,
Berkowitz_prb_model, Oscar_prb}).

Since maghemite core and shell properties have a different
pressure response, one can expect that maghemite particles with
different geometry have a different behaviour with pressure,
allowing a better insight on the magnetic properties of core and
shell. Hollow maghemite NPs are an exotic and interesting system
where the relevance of surface is
enhanced\cite{doi:10.1021/ja072574a, PhysRevB.79.094419}.
Accordingly, we investigate the effect of pressure in hollow
maghemite NPs and we compare this effect with that observed in
solid maghemite NPs obtained by polymeric-assisted synthesis and
non-aqueous routes.

\section{\label{sec:exp}Experimental}

Three different samples were synthesized and studied: a sample
composed of hollow iron oxide NPs capped by oleylamine, a sample
composed of solid iron oxide NPs capped by oleic acid and a sample
composed of solid iron oxide NPs dispersed in a polymer and
forming a composite.

Hollow iron oxide NPs were obtained by the nanoscale Kirkendall
effect following a previously reported
procedure\cite{doi:10.1021/ja072574a}. Briefly, iron NPs were
obtained by decomposition of iron pentacarbonyl in organic
solvents containing amines. 10 ml of octadecene (C$_{18}$H$_{38}$)
containing 0.67~mmol of oleylamine were heated inside a three-neck
flask to 60~$^\circ$C under vacuum for 30~min. While keeping the
solution under argon, the temperature was raised to 200~$^\circ$C.
A precursor solution of 0.4~ml of Fe(CO)$_5$ in 2~ml of octadecene
was prepared separately under Ar. This was rapidly injected
through a septum into the hot surfactant solution under vigorous
stirring. The resulting solution was reacted for 20 min.
Afterwards, to oxidize the formed iron NPs, 20~ml/min of a 20\%
oxygen mixture in argon were flowed through the heated flask over
2~h. 

Solid $\gamma$-Fe$_2$O$_3$ NPs were obtained by decomposing iron
pentacarbonyl in octadecene in the presence of an excess of oleic
acid. The presence of equivalent or excess amounts of oleic acid
in the precursor solution initially results in the formation of
iron oleate, which decomposes directly into iron oxide. In a
typical synthesis, a mixture of 10 ml of octadecene and 2 ml of
oleic acid were heated inside a three-neck flask to 60 $^\circ$C
under vacuum for 30 min. While keeping the solution under argon,
the temperature was raised to 280 $^\circ$C. At this temperature
0.4 ml of Fe(CO)$_5$ in 2 ml of octadecene were rapidly injected
through a septum. The resulting solution was reacted for 20 min.
The particle size was tuned by changing reaction time and the
concentration of iron carbonyl injected.

 Maghemite/polymer
nanocomposites were prepared from iron/poly(4-vinylpyridine) (PVP)
precursor compounds, following the procedure reported in
Ref.\cite{Angel_Comp}. The precursor was prepared by mixing a
water:acetone solution of PVP with a RbBr-FeBr$_2$-FeBr$_3$ stock
solution. The solution was evaporated at 40~$^\circ$C to obtain a
solid film. The precursor film was immersed in a 1 M NaOH solution
for 1~h, washed with water and dried at 60~$^\circ$C to obtain a
polymer/NPs composite. Finally, the composite was annealed at
200~$^\circ$C for 24~h. In the following, the three studied
samples are identified as hollow, solid and polymer-grown.

The geometry, size and crystallographic structure of the NPs were
characterized by transmission electron microscopy (TEM) and
high-resolution TEM (HRTEM) using a Jeol 2010F field emission gun
microscope with a 0.19~nm point to point resolution.

X-ray diffraction (XRD) measurements were performed at room
temperature as a function of pressure up to 30~kbar using a
cylindrical imaging plate diffractometer (Rigaku Co.) at the
Photon Factory of the Institute of Materials Structure Science at
the High Energy Accelerator Research Organization (KEK)
\cite{SynchrotronPressure}. The wavelength of the incident X-ray
was $\lambda=0.68850(2)$~\AA. 
Pressure was applied using a diamond anvil cell, which consisted
of two diamond anvils with flat tips of diameter 0.8~mm and a
0.3-mm-thick CuBe gasket. The pressure was calibrated by the ruby
fluorescence method\cite{PressureMethodRubi}. The maghemite NPs
and a few ruby crystals were held along with a
pressure-transmitting medium (fluorinated oil, FC77, Sumitomo 3M
Co., Ltd.) in a sample cavity of diameter 0.4~mm located at the
center of the CuBe gasket. The analysis of the diffraction
patterns was performed by Rietveld refinement using the FullProf
package \cite{FullProf}. The size effects were treated with the
integral breadth method using the Voigt model for both the
instrumental and intrinsic diffraction peak shape considering a
Thompson-Cox-Hastings pseudo-Voigt convoluted with Axial
divergence asymmetry function to describe the peak shape. The
contribution of the finite size of the NPs crystallites to the
peaks broadening was taken into account by an isotropic model
yielding an average apparent size.

For the magnetic measurements, the pressure was generated by a
piston-cylinder-type CuBe pressure cell that was designed to be
inserted into a commercial superconducting quantum interference
device (SQUID) magnetometer (Quantum Design,
MPMS)\cite{MitoPressureCell}. The maghemite NPs were held in the
Teflon cell, which was installed in the pressure cell along with
the pressure-transmitting medium (Apiezon-J oil) and a few pieces
of superconductor tin used as a manometer. The pressure at
liquid-helium temperature was estimated by the shift in the
superconducting transition temperature of
tin\cite{PressureMethod}. The ac magnetic response was measured as
a function of the temperature ($T$) and the frequency of the ac
field ($f$). Under an ac field ($H_{ac}$) of 4.0~Oe, the in-phase
($\chi$') and out-of-phase ($\chi$'') components of a series of
first-order harmonic components, $M1\omega/h=\chi1\omega$, were
detected from the Fourier transform of the SQUID voltage, which
was measured after modification of the phase delay due to the eddy
current of CuBe at each frequency.

\section{\label{sec:res}Results and discussion}
%

\begin{figure}[htb!]
\begin{center}\includegraphics[width=0.8\columnwidth]{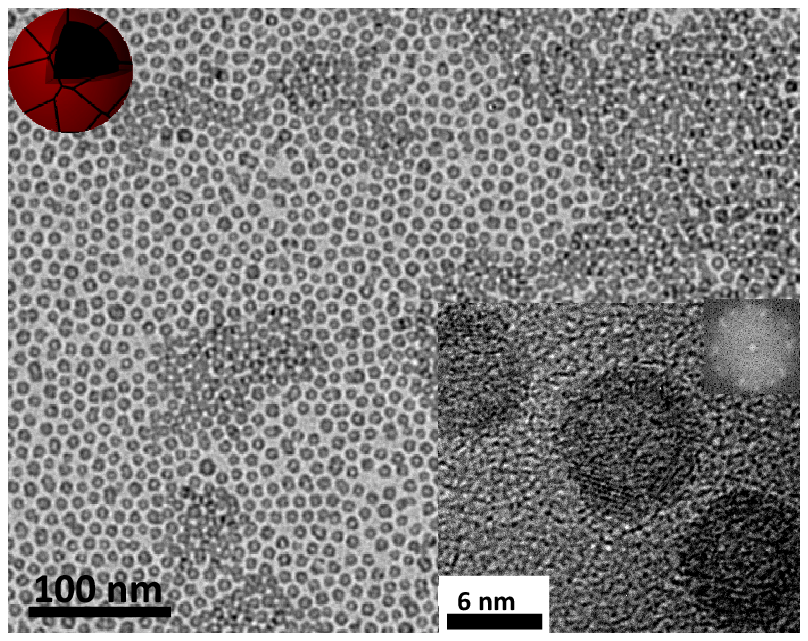}   
\end{center}
\caption{\label{Fig:TEM} Typical TEM micrograph of the hollow
maghemite NPs with 8~nm average size. Inset micrograph corresponds
to a high resolution image of the same NPs and corresponding
Fourier-transform. Top-left inset cartoon depicts the hollow
polydomain structure of a NP.}
\end{figure}

TEM micrographs of the hollow sample show the expected geometry of
core/shell hollow/solid NPs with an average diameter of about 8~nm
and a low size dispersion (Fig.\ref{Fig:TEM}). The iron oxide
shell has about 3~nm, being polycrystalline, as seen in the HRTEM
micrographs and corresponding Fourier transform (Fig.
\ref{Fig:TEM}, right inset). Accordingly, the NP structure can be
depicted as a tectonic crust, as shown in Fig. \ref{Fig:TEM},
top-left inset.

%
%
%

\begin{figure}[htb!]
\begin{center}\includegraphics[width=0.9\columnwidth]{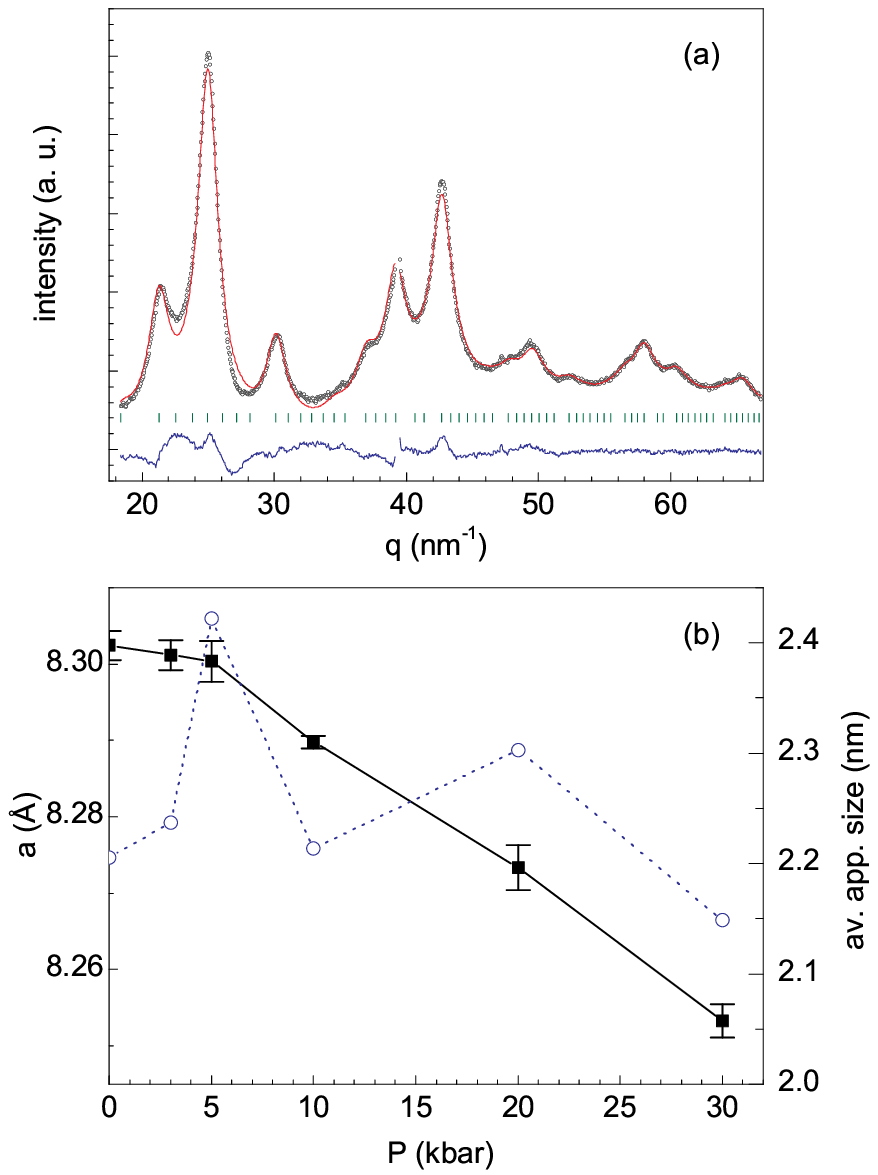}
\end{center}
\caption{\label{Fig:XRD} (color online) (a)  Room temperature and
ambient pressure X-ray diffraction (XRD) pattern of the hollow
maghemite NPs. Continuous (red) line corresponds to Rietveld
refinement of a spinel as described in the text, vertical lines
represent the position of allowed Bragg peaks, while horizontal
(blue) line represents the fit residues. (b) pressure dependence
of the cell parameter $a$ (left scale, full symbols) and average
apparent size (right scale, open symbols); solid lines are eye
guides.}
\end{figure}

 XRD patterns of the hollow sample show the existence of NPs with a spinel
 structure consistent with magnetite/maghemite [Fig.\ref{Fig:XRD}(a)]. The patterns can be well
 reproduced by considering the $P 4_3 3 2$ space group and a peak
 broadening due to finite size effects. Fits with similar quality
 are obtained when considering different Fe/O stoichiometry
 although the best fit is obtained with a stoichiometry closer to
 magnetite than to maghemite. This result should be taken
 carefully, since the background is ill-defined making difficult a
 proper determination of the relative intensity of the peaks. In
 fact, previous spectroscopy studies suggest that the iron oxide
 is maghemite rather than magnetite\cite{doi:10.1021/ja072574a}. The
 contribution to the peak broadening due to strain is negligible
 compared to that of size.
 The average apparent size at room pressure is $\sim 2.2$ nm [Fig.\ref{Fig:XRD}(b)], in
 good agreement with the $\sim 3$ nm crystalline domains observed
 by HRTEM.
The cell parameter decreases monotonically with pressure, whereas
the average apparent size has no defined trend having values in
the 2.1 to 2.4~nm range (which is probably close to its error
bar).

%

\begin{figure}[htb!]
\begin{center}\includegraphics[width=0.9\columnwidth]{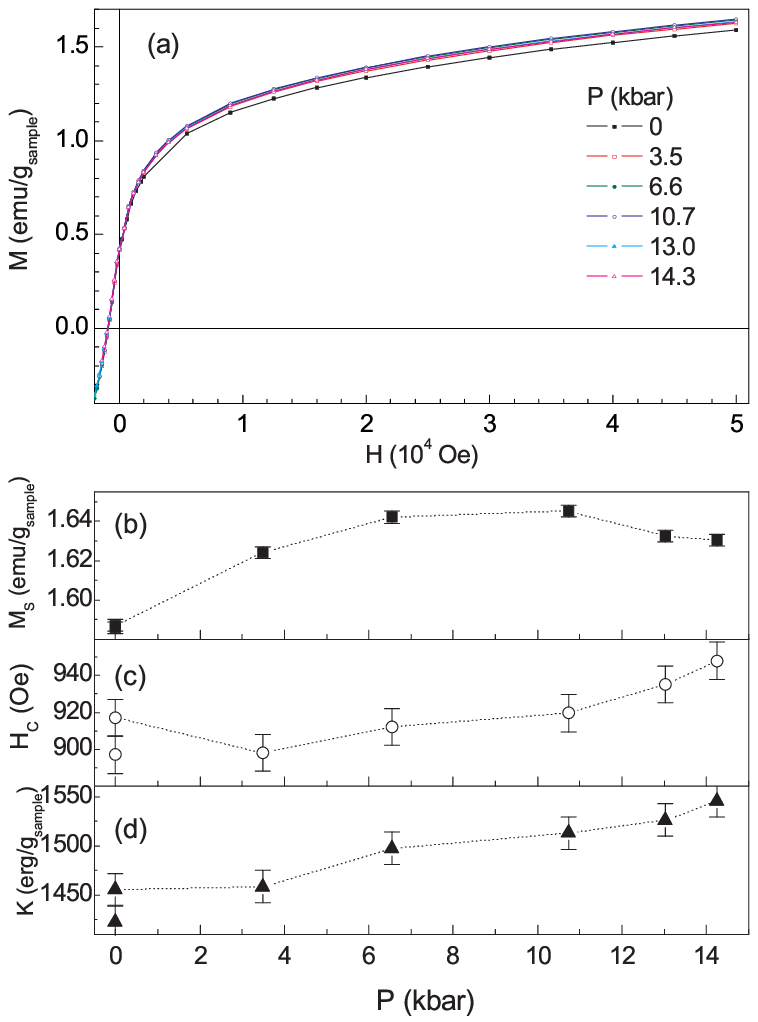}
\end{center}
\caption{\label{Fig:MH}(color online) (a) Field dependence of the
magnetization of hollow NPs measured at decreasing fields after
zero-field cooling and obtained at selected pressures and $T=5$ K.
Pressure dependence of the (b) magnetization at high field
($5\times 10^4$~Oe) $M_S$ obtained at $T=5$ K, (c) coercive field
$H_C$ and (d) effective anisotropy constant $K_{eff}$. Lines are
eye guides for data obtained at increasing pressures. The values
obtained at ambient pressure after pressure release are shown as
isolated symbols. Error bars in panel (b) and (c) are estimations
based on the standard deviation of consecutive measurements of
magnetization while error bars in panel (d) are obtained by
propagation of error.}
\end{figure}

At low temperature, magnetization shows hysteresis with field
[Fig.\ref{Fig:MH}(a)]. The coercive field $H_C$ and the
magnetization at the maximum field used in the experiment (denoted
as $M_S$) are pressure dependent, increasing and decreasing with
pressure, respectively [Fig.\ref{Fig:MH}(b) and (c)]. Taking into
account these two dependencies it is possible to evaluate the
pressure dependence of the effective anisotropy constant
$K_{eff}$, since $K_{eff}\propto H_C M_S$. Despite the opposite
trends of $H_C$ and $M_S$, $K_{eff}$ increases with pressure,
anticipating a pressure dependence of the anisotropy energy
barrier.

\begin{figure}[htb!]
\begin{center}\includegraphics[width=0.9\columnwidth]{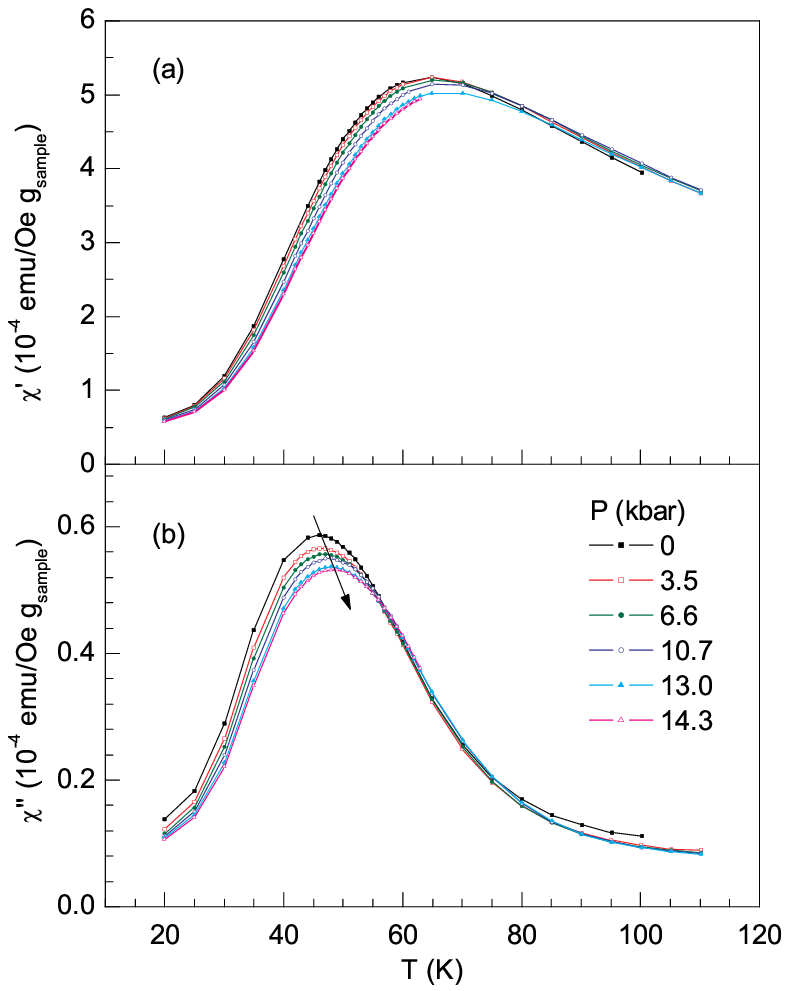}
\end{center}
\caption{\label{Fig:Sus} (color online) In-phase (a) and
out-of-phase (b) components of the ac susceptibility of the hollow
NPs obtained as a function of temperature for selected applied
pressures with an excitation ac field of 1~Hz. Arrow denotes the
trend of the maximum position with the increase of pressure.}
\end{figure}

\begin{figure}[htb!]
\begin{center}\includegraphics[width=0.9\columnwidth]{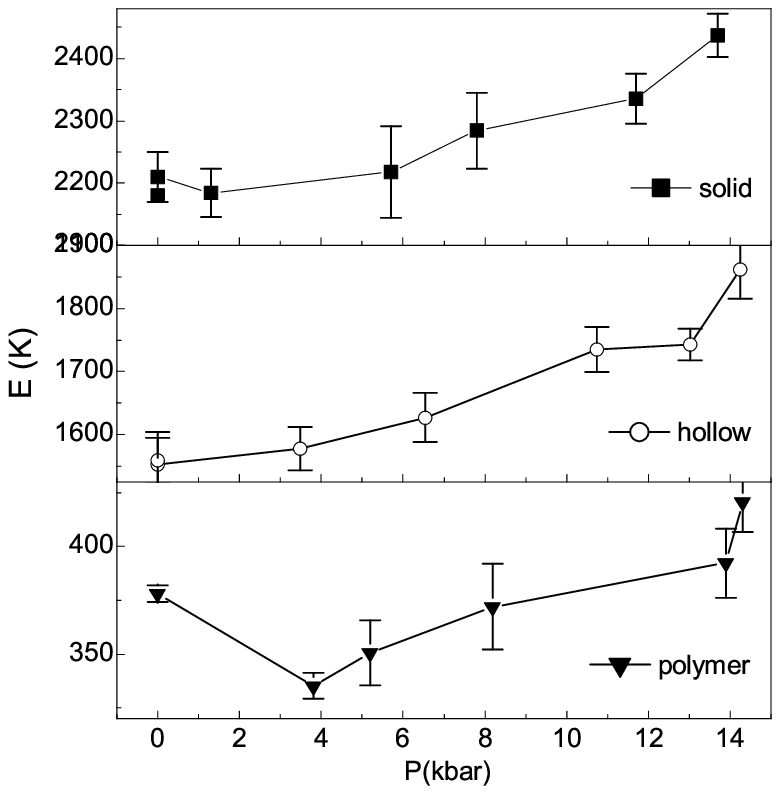}  
\end{center}
\caption{\label{Fig:EP} Pressure dependence of the anisotropy
energy barrier $E$ for the solid, hollow and polymer-grown
maghemite NPs. Solid lines are eye guides for data obtained at
increasing pressures. The values obtained at ambient pressure
after pressure release are shown as isolated symbols. Error bars
were estimated based on linear fits to Arrhenius plots. }
\end{figure}

The temperature dependence of the ac susceptibility at room
pressure of the hollow and solid sample shows the characteristic
features of superparamagnetic NPs with a distribution of energy
barriers undergoing an unblocking process as temperature increases
from 20 to 100 K, showing a frequency ($f$) dependent maximum with
temperature (blocking temperature $T_B$) at around $T_B=45$~K
(Fig.\ref{Fig:Sus}). With the increase of pressure, $T_B$ at a
fixed frequency increases to higher temperatures. 
%
%
%
%
%
%
%
At a given pressure, $T_B(f)$ follows a N\'{e}el-Arrhenius
relation, $\tau_m=\tau_0 \exp(E/k_B T_B$) as usually found in
superparamagnetic NPs\cite{Neel_arrhenius, Brown}. Here $\tau_0$
is a microscopic characteristic time, $\tau_m$ is the
characteristic measurement time equal to $1/(2\pi f)$ and $E$ is
the anisotropy energy barrier, usually expressed as the product
between $K_{eff}$ and the NPs average volume $V$. From the
N\'{e}el-Arrhenius relation, the pressure dependence of $E$ can be
estimated (Fig.\ref{Fig:EP}). Qualitatively, the pressure
dependence of $E$ is similar in the solid, hollow and
polymer-grown maghemite NPs ($E_{solid}$, $E_{hollow}$ and
$E_{polymer}$), increasing with pressure in the studied range. The
increase of $E_{hollow}$ is in accordance with the increase of
$K_{eff}$ determined from the magnetization results.

\begin{figure}[htb!]
\begin{center}\includegraphics[width=0.9\columnwidth]{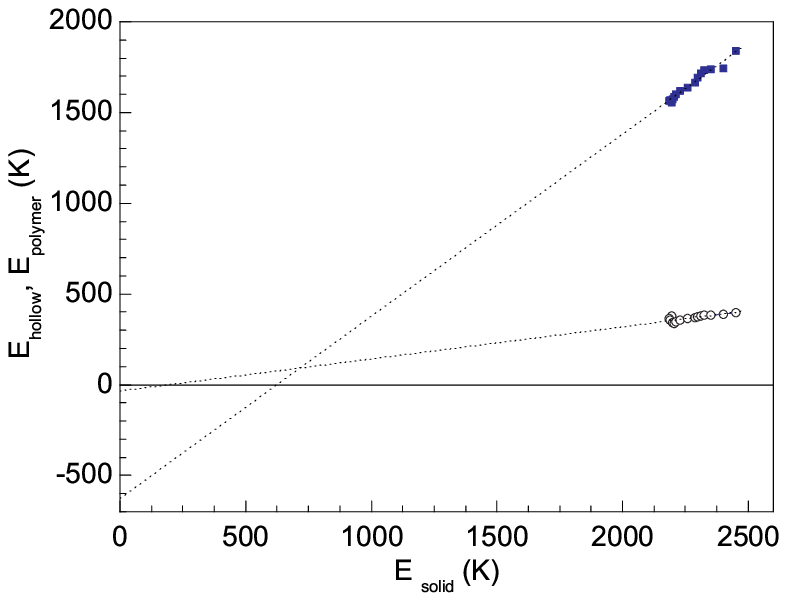}
\end{center}
\caption{\label{Fig:EvsE}(color online) Relation between the
anisotropy energy barrier $E$ of the solid and those of the hollow
and polymer-embedded maghemite NPs. Dotted lines represent linear
fit and low energy extrapolation.}
\end{figure}

A better insight on the relation between $E$ obtained for the
three studied samples is obtained by plotting two of them as a
function of the third (Fig.\ref{Fig:EvsE}). Since $E$ of the three
samples was estimated at different pressure values, a simple
interpolation procedure was applied. In the plot of
Fig.\ref{Fig:EvsE}, the similitude between the $E$ pressure
dependence is apparent as a linear dependence between $E$ of the
solid NPs and those of the hollow and polymer-grown NPs.
Interestingly, while $E_{polymer}$ is simply proportional to
$E_{solid}$ with the linear extrapolation crossing the (0,0)
point, $E_{hollow}$ is proportional to $E_{solid}$ with the linear
extrapolation crossing the x-axis at a positive value, such that

\begin{eqnarray}
 E_{hollow}= E_{solid}-627\quad(K)    \nonumber \\
E_{polymer}=0.176 E_{solid}
\end{eqnarray}
This means that the pressure dependence of the polymer-grown NPs
and that of the solid maghemite NPs has the same physical origin,
differing only by a constant term reflecting the different $E$
value of both samples at ambient pressure, associated with their
different $V$ and $K_{eff}$. On the other hand, the slope of the
$E_{hollow}$ $vs.$ $E_{solid}$ dependence is quite close to 1
while when $E_{hollow}$ extrapolates to zero, $E_{solid}$ has
still a non-zero contribution of the order of 627~K. In a first
approximation, this can be regarded as the $E_{solid}$ having two
components; one component displaying a behavior similar to that of
$E_{hollow}$ (the linear contribution) and a second component
which is absent in $E_{hollow}$ (the non-zero contribution at
$E_{hollow}=0$). By geometrical arguments, the component common to
both solid and hollow sample is the surface, while the second
component present in the solid sample and absent in the hollow one
is the core. This suggests that $E$ associated with the surface
has the most relevant pressure dependence while $E$ associated
with the core has a relevant contribution at ambient pressure,
being almost pressure independent.

\section{Conclusions}
In summary, it was shown that the anisotropy energy of solid
maghemite NPs prepared by different routes of synthesis have a
similar pressure dependence, while the anisotropy energy of solid
and hollow maghemite NPs show different pressure dependence. This
difference is due to the different geometry of the NPs and with
the larger pressure response of the shell.

\section*{Acknowledgments}

The Aveiro- Barcelona collaboration has been supported by the
Integrated Spanish-Portuguese Action under the Grant No.
AIB2010PT- 00099. The Aveiro-Zaragoza collaboration has been
supported by the Integrated Spanish-Portuguese Action PT2009-0131.
The work in Zaragoza has been supported by the research grants
MAT2011-27233-C02-02, MAT2011-25991 and CONSOLIDER CSD2007-00010
from the Ministry of Education. The financial support of the CSIC
and Spanish Ministerio de Ciencia e Innovaci\'{o}n (PI201060E013)
is also acknowledged. The work in Japan was supported by a
Grant-in-Aid for Scientific Research (C) (No. 23550158) from the
Ministry of Education, Culture, Sports, Science and Technology
(MEXT), Japan. \`{O}. I. and A. L. acknowledge funding of the
Spanish MICINN through Grant No. MAT2009-08667 and No.
CSD2006-00012, and Catalan DIUE through project No. 2009SGR856. N.
J. O. S. acknowledges FCT for Ciencia 2008 program.

\section*{Bibliography}

\end{document}